\documentclass[journal]{IEEEtran}
\usepackage{comment}

\usepackage{ifpdf}

\usepackage{algorithmicx}
\usepackage{times}
\usepackage{fancyhdr,graphicx,amsmath,amssymb}
\usepackage[ruled,vlined]{algorithm2e}
\usepackage{cite}
\include{pythonlisting}
\makeatletter
\def\BState{\State\hskip-\ALG@thistlm}
\makeatother

\usepackage{cite}
\usepackage{amsmath}
\usepackage{amssymb}
\usepackage{flexisym}
\usepackage{array}
\newcolumntype{C}[1]{>{\centering\let\newline\\\arraybackslash\hspace{0pt}}m{#1}}

\usepackage{color}
\definecolor{dblue}{rgb}{0,0,0.8}

\usepackage{xcolor}

\usepackage{multirow}

\usepackage{subfig}
\usepackage{graphicx}
\usepackage{fixltx2e}

\usepackage{stfloats}
\usepackage{subfig}

\usepackage[utf8]{inputenc}

\usepackage{amssymb}

\usepackage{nomencl}
\makenomenclature
\usepackage{etoolbox}
\usepackage{hyperref}

\usepackage{url}

\usepackage{mhchem}

\usepackage{nomencl}
\makenomenclature

\usepackage[justification=centering]{caption}

\usepackage{nomencl}
\makenomenclature

\usepackage{hyperref}

\begin{document}
%


\title{Proactive Rolling-Horizon based Scheduling of Hydrogen Systems for Resilient Power Grids}



%
%
%

\author{Hamed~Haggi,~\IEEEmembership{Student Member,~IEEE,},
        Wei~Sun,~\IEEEmembership{Member,~IEEE},
        James M. Fenton, and Paul Brooker
\thanks{This work is supported by U.S. Department of Energy's award under grant DE-EE0008851. H. Haggi and W. Sun are with the Department of Electrical and Computer Engineering, University of Central Florida, Orlando, FL 32816 USA (e-mail: hamed.haggi@knights.ucf.edu, sun@ucf.edu). J. M. Fenton is with Florida Solar Energy Center, University of Central Florida, Cocoa, FL 32922 USA (jfenton@fsec.ucf.edu) and P. Brooker is with Orlando Utilities Commission, Orlando, FL 32839 USA (PBrooker@ouc.com).}}

\maketitle

\begin{abstract}
Deploying distributed energy resources (DERs) and other smart grid technologies have increased the complexity of power grids and made them more vulnerable to natural disasters and cyber-physical-human (CPH) threats. To deal with these extreme events, proactive plans are required by utilities to minimize the damages caused by CPH threats. This paper proposes a proactive rolling-horizon-based scheme for resilience-oriented operation of hydrogen (H2) systems in integrated distribution and transmission networks. The proposed framework is a bi-level model in which the upper-level is focused on distribution system operation in both normal and emergency operation modes, and the lower-level problem accounts for the transmission network operation. Two preeminent aspects of H2 systems are considered in this paper, 1) to show the flexibility of H2 systems, capacity-based demand response signals are considered for electrolyzers, stationary fuel cell (FC) units, and H2 storage tanks are considered in both normal and emergency operation modes; 2) unlike the batteries which can only charge and discharge energy based on maximum duration times and power ratings, H2 systems can be considered as the flexible long-term energy storage by storing H2 for days and supplying power to FC in the case of \textit{N-m} outages lasting for more than 10 hours. Moreover, H2 production cost based on water electrolysis and storage costs is calculated. Simulation results demonstrate that utilities can improve the system-level resilience using H2 systems as long-term backup power resources. 
\end{abstract}

%
\IEEEpeerreviewmaketitle

\section{Introduction}
\IEEEPARstart{P}{roliferation} of distributed energy resources (DERs) and smart grid technologies, have driven the power systems more complex and vulnerable to cyber-physical-human (CPH) threats and natural disasters  \cite{haggi2019review}. These threats can significantly affect the operation of power systems, especially distribution networks due to the radial topology, limited backup power, and overhead line outages \cite{khazeiynasab2020resilience}. To increase the system resilience and minimize the impact of these high-impact low-probability events on power systems in both normal and emergency operation, proactive schemes must be considered by utilities, such as using flexible DERs (e.g. energy storage, microgrids, etc.), to provide backup power sources for delivering power especially to the critical loads.\par

Generally, the power systems resilience can be improved by proactive plans prior and after a disruption, survivability analysis, and restorative schemes. Since the scope of this paper is proactive scheduling plans, the authors only reviewed the related proactive and survivability research efforts in resilience enhancement topic (more details regarding the post-event and restoration part can be found in our previous work \cite{haggi2019review}). For instance, in \cite{nguyen2020preparatory}, a stochastic model was developed for preparatory operation of distributed energy storage systems prior to hurricanes. Additionally, post-event decisions were also considered to enhance the resilience by restoring the critical loads. Authors of \cite{gholami2017proactive} proposed a two-stage adaptive robust approach to enhance the resilience by minimizing the damaging consequences using microgrids. In \cite{hussain2018proactive}, both normal and emergency operation considering resilience cut for battery energy storage and microgrids were considered to improve the resilience of system. Moreover, uncertainties of loads and renewable generation were considered in their model. A proactive linearized plan was proposed in \cite{amirioun2017resilience} using microgrids to cope with windstorms. Network reconfiguration, demand side management, etc. were considered to prevent load curtailment. Moreover, these authors' work was extended by considering both electric and gas networks \cite{amirioun2018resilience}. 

System operators can schedule their assets, especially DERs, to prevent damages which may be caused by natural disasters or CPH threats. In the recent years, H2 energy is of great interest of researchers due to its environmental and technical merits in both power and transportation networks. To this end, different applications of H2 energy have been mainly focused on, 1) power generation by FC units for grid balancing purposes, 2) fuel for transportation sector by supplying the H2 demand of heavy duty trucks, fuel cell electric vehicles (FCEVs), and aviation sector, 3) feedstock for industry such as ammonia production, etc. \cite{kovavc2021hydrogen}. For instance, authors of \cite{khani2019supervisory} proposed a model in which distributed H2 fueling stations participate in reserve market based on their free capacity to increase the profit. In \cite{el2018hydrogen}, H2 fueling stations (including electrolyzer and storage tank) were optimally scheduled considering the demand response signals with the aim of maximizing the total profit of private owner of these distributed fueling stations. 
A design for onsite H2 production was proposed in \cite{xiao2010hydrogen} with the goal of minimizing the operational cost as well as supplying the H2 demand of fuel cell vehicles. A techno-economic feasibility analysis including H2 energy storage systems was investigated in \cite{eypasch2017model}. A decentralized game theory-based local market for H2 and electricity trading considering the H2 vehicles demand was investigated in \cite{xiao2017local}. Authors of \cite{wu2020cooperative} proposed a distributed coordinated operation framework for wind and H2 fueling stations considering uncertainty of wind and electricity price. 

Considering the aforementioned discussion, research efforts are mainly focused on normal operation scheduling of H2 systems \cite{kovavc2021hydrogen}-\cite{wu2020cooperative}, and the techno-economic merit of H2 systems in enhancing the grid resilience has not been investigated yet. However, unlike the battery energy storages which can only store, charge, and discharge energy based on their maximum energy rating (maximum duration time), H2 systems can be considered as long-term energy storages to produce H2 by electrolyzers, store it in the tank, then convert it to power with FC units and inject power into the grid with the maximum FC capacity for longer period of time (e.g. days, months, etc.). In addition,the coordinated operation scheduling of distribution and transmission networks has been only focused on normal operation and planning context \cite{el2021coordinated}-\cite{hassan2018energy}. However, proactive scheduling of DERs in integrated transmission and distribution systems with the aim of resilience enhancement has not been investigated. Additionally, previous proactive scheduling frameworks only addressed the distribution network operation without considering the benefits of selling or purchasing power from transmission networks \cite{nguyen2020preparatory}-\cite{amirioun2018resilience}. Therefore, this paper extends authors' previous work presented in \cite{haggi2021proactive} by developing a bi-level framework for resilient scheduling of H2 systems in integrated distribution and transmission networks. The major contributions of this paper compared to \cite{haggi2021proactive} are:

\begin{itemize}
  \item A bi-level resilience-oriented framework considering the coordinated operation of distribution and transmission networks with the focus on scheduling the H2 systems (including both H2 refueling station and long-term energy storage system), in both normal and emergency operation mode is proposed. The upper-level and lower-level problems are focused on distribution network managed by distribution system operator (DSO) and transmission network managed by transmission system operator (TSO), respectively. To efficiently solve the bi-level problem, duality theory is deployed to recast it as a single-level equivalent problem.
  \item Capacity-based demand response (CBDR) signals are considered in pre-event operation. In normal operation, H2 systems can follow the signals imposed by DSO and assist the grid by acting as load (operating the electrolyzers) or generation unit (operating FC units). For the emergency operation preparation, as soon as having the access to extreme event time based on the forecast, DSO sends signals to H2 systems to fill their storage tanks and be prepared for long-duration outages. 
  \item Rolling horizon approach is deployed to limit the access of DSO to the perfect forecast of extreme event time (by providing only the next-day forecast), CBDR signals, and other information of renewables. Additionally, rolling-based operation can show the capability of H2 systems as fast-response DERs which provides more realistic results.
  \item Water electrolysis and storage costs are calculated for defining the selling price of H2 to FCEVs and providing a realistic revenue for DSO. It should be mentioned that H2 production cost is calculated based on real capacity factor (CF) of electrolyzers, distributional locational marginal price (DLMP), etc., which is more realistic due to the consideration of integrated operation of distribution and transmission networks energy price, congestion, power loss, and voltage regulation costs. 
\end{itemize}\par
The rest of the paper is organized as follows. Section \ref{framework} introduces the framework of this paper. Section \ref{formulations} presents the problem formulation. Section \ref{results} presents numerical results and analysis, and finally Section \ref{conclusion} concludes the paper and presents future work directions. More details on the linearization of the problem formulation is presented in Appendix.

\vspace{-0.2cm}
\section{Proposed Framework}\label{framework}
\begin{figure}
\centering
\footnotesize
\captionsetup{justification=raggedright,singlelinecheck=false,font={footnotesize}}
	\includegraphics[width=3.5in,]{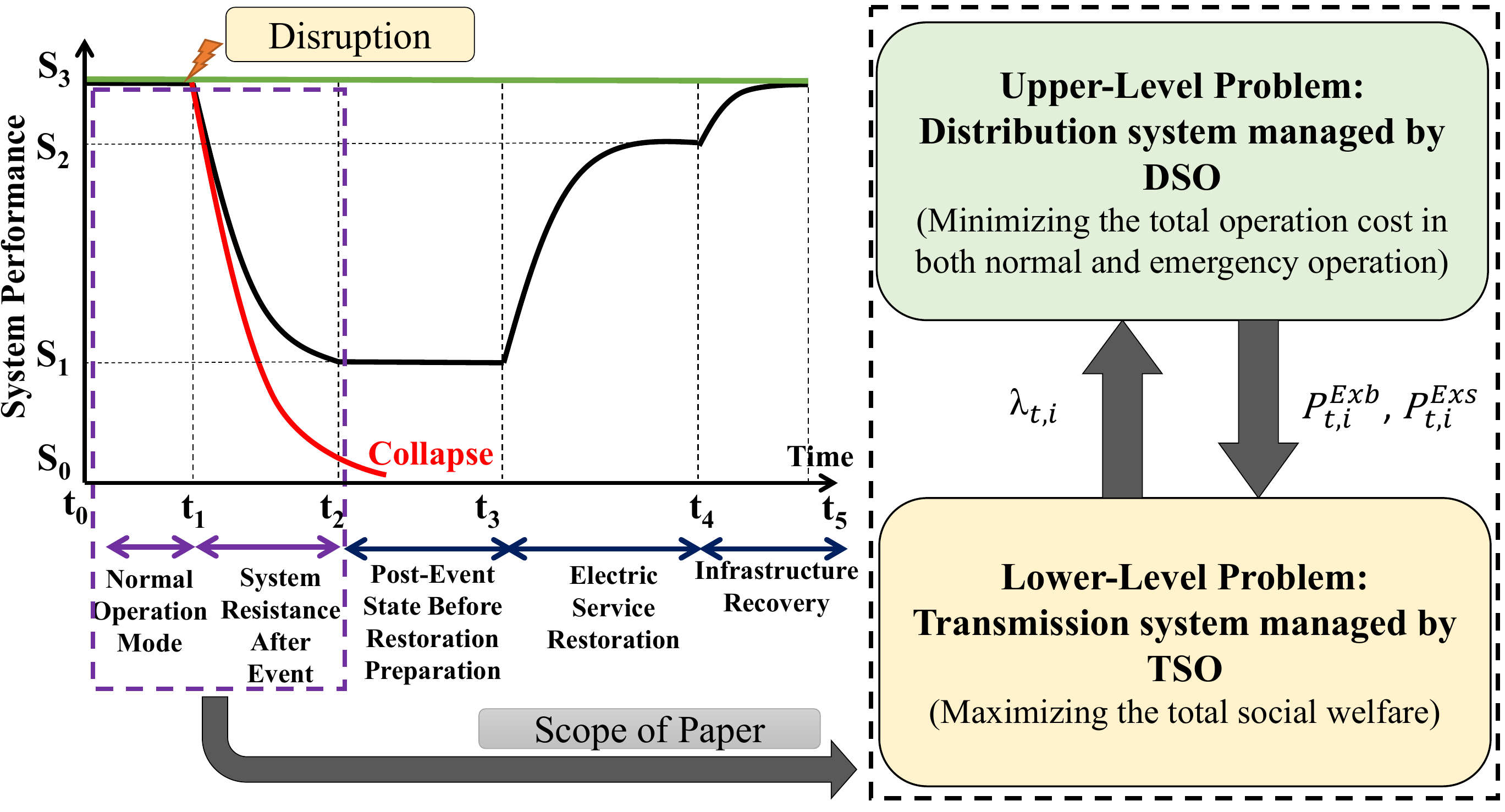}
	\caption{The proposed bi-level resilience-oriented framework for integrated transmission and distribution networks.}
    \label{scope}
\end{figure}

The proposed bi-level resilience-oriented framework with the focus on proactive scheduling of integrated transmission and distribution networks is presented in Fig. \ref{scope}. It shows that the scope of this paper is proactive scheduling and survivability analysis for the normal and emergency operation, respectively. The upper-level (UL) problem minimizes the total operation cost of all assets in both normal and emergency operation modes from DSO's perspective. On the other hand, the lower-level (LL) problem maximizes the total social welfare of the wholesale market managed by TSO. In this framework, distribution network (DN) is connected to transmission network (TN) via single root bus. It should be mentioned that, DSO participates in wholesale market by submitting the offers $P_{t,i}^{Exs}$ and bids $P_{t,i}^{Exb}$ for selling and purchasing power, respectively. On the other hand, TSO defines the locational marginal price (LMPs) which is defined as $\lambda_{t,i}$.\par

\begin{figure}
\centering
\footnotesize
\captionsetup{justification=raggedright,singlelinecheck=false,font={footnotesize}}
	\includegraphics[width=3.5in, height=2in]{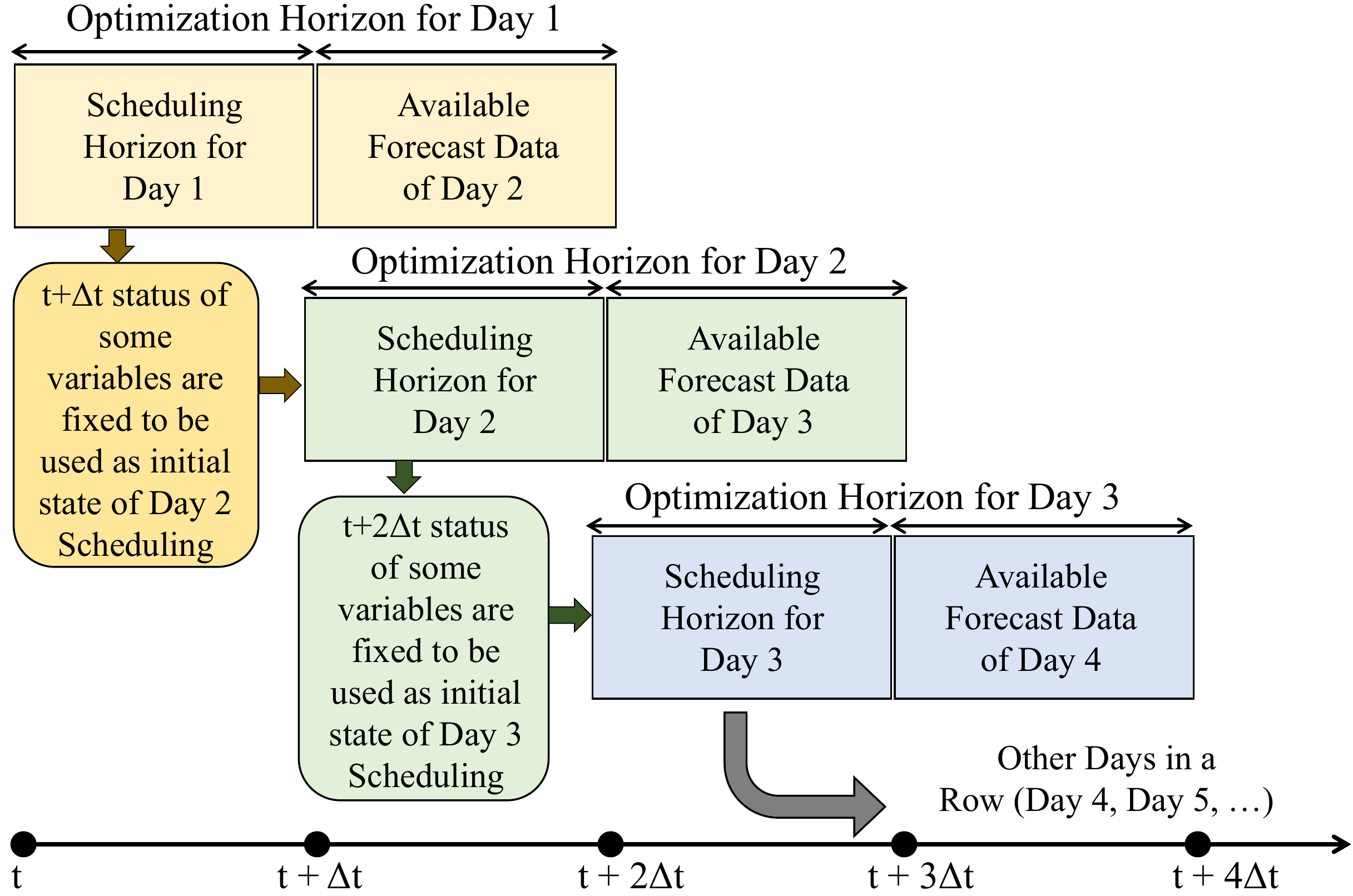}
	\caption{Rolling-horizon based approach for resilient day-ahead scheduling.}
    \label{rolling}
\end{figure}

In this resilience-oriented framework, a vertically integrated DN is considered in a way that DSO operates utility-operated photovoltaic (PV) units and natural gas power plants. DSO must supply the power to H2 systems, including electrolyzers, storage tanks, and stationary fuel cell (FC) units. In the normal operation, distributed H2 systems are scheduled to exploit renewable energy resources and minimize the total operation cost and energy not supplied (based on load priority). The H2 production cost, consisting of water electrolysis cost and storage cost, is calculated based on the distributional locational marginal prices (DLMPs) considering the LMP prices of transmission network. Additionally, CBDR signals are incorporated into the optimization problem to demonstrate the flexibility of H2 systems. Prior to an emergency operation mode, DSO sends emergency CBDR signals to H2 systems in order to fill their storage tank and be prepared for post-event times. This results in maximizing the survivability by using the stored H2 for stationary FC consumption and consequently the resilience improvement. Moreover, the rolling horizon approach, as presented in Fig. \ref{rolling}, is applied to the bi-level framework, in order to address the challenge from the unavailable perfect forecasts for system operators. With this model, DSO does not know the exact time of disruption, output power of renewable energy resources, transportation sector demand, and CBDR signals for scheduling; However, DSO only has the access to next 24-hour forecasts (total rolling horizon period is 48 hours). At each time period, the final status of DGs, mass of H2 in the tank state, etc., will be fixed as initial condition for the next rolling horizon-based scheduling. This results in more realistic results in the case of major disruption. In this paper, the optimization horizon is the same for both DSO and TSO.\par

\vspace{-0.2cm}
\section{Problem Formulation}\label{formulations}
The problem formulation of the proposed bi-level resilience-constrained problem (RCP) is presented in this section. The proposed RCP model is formulated as a mixed integer quadratic constrained program (MIQCP). Given a network, $(\mathcal{N},\mathcal{L})$, where $\mathcal{N}$ and $\mathcal{L}$ are the set of nodes and lines indexed by $i$ and $l$. For DN and transmission network, the sets of nodes and lines are $\mathcal{N_D}$ and $\mathcal{L_D}$, $\mathcal{N_T}$ and $\mathcal{L_T}$, respectively. The root node of DN is connected to the bus two of TN. $\mathcal{T}$ represents the set of time steps indexed by $t$.  

\vspace{-0.3cm}
\subsection{Upper-Level Problem Formulation: DSO Perspective}
\subsubsection{Objective Function of UL Problem} The objective function of the UL is to minimize the total operation cost in both normal and emergency conditions. For the sake of brevity, the problem formulation is not divided into normal and emergency operation modes. However, prior to the major disruption, all equations are valid except for load shedding terms, which should be removed or considered as zero in the equations.  
\begin{equation}\label{ULOF}
\begin{split}
\text{min.} \sum_{t=1}^{T}\Bigg\{&  \lambda_{t,i} \;.\; P_{t,i}^{Exb} -\beta_{t,i}\;.\;P_{t,i}^{Exs} + \sum_{i=1}^{N_G} C^{DG}_{t,i} + \sum_{i=1}^{N_G} C^{SU}_{t,i} + \\& \sum_{i=1}^{N_G} C^{SD}_{t,i} + \sum_{i=1}^{N_{PV}} C^{PV}_{t,i} + \sum_{i=1}^{N} C^{Load,Shd}_{t,i}  \Bigg\}
\end{split}
\end{equation}
In (\ref{ULOF}), the first term refers to the purchasing power from the wholesale market. The second term refers to selling power from the distribution network to the wholesale market. The third, forth, and fifth terms refer to operating cost, startup, and shutdown cost of DGs, respectively. The next two terms refer to operational cost of utility operated PV units and cost of energy not supplied based on the load priority (e.g. critical, moderately-critical, and non-critical grid load). It should be noted that the costs associated with H2 systems are not taken into account and DSO only schedules its system demand. In this paper, both DSO and H2 system owners perform the cost benefit analysis separately, and exchange energy with power purchase agreement price. Moreover, the cost of FCEVs demand curtailment is not considered in the objective function since the value of loss of EV load is negligible compared to loss of grid load.\par

\subsubsection{Operational Constraints of DG Units}
The operational constraints for utility-operated DGs are shown in (\ref{DGcost})-(\ref{DGramp}). Equations (\ref{DGcost})-(\ref{QDGlimit}) show the operational cost and active/reactive power output limits of DGs, respectively. Equation (\ref{DGSOCP}) expresses the reactive power support limit based on active power flow. Moreover, equations (\ref{DGSU1})-(\ref{DGSD2}) show the startup and shutdown cost of DGs, respectively. Finally, equation (\ref{DGramp}) presents the ramping up and down limits of DGs. 
\begin{equation}\label{DGcost}
C^{DG}_{t,i} = x^{DG}_{t,i}.\; b^{DG}+ k^{DG}.\;P^{DG}_{t,i}
\end{equation}
\begin{equation}\label{PDGlimit}
P^{DG,min}_{i}.\;x^{DG}_{t,i} \le P^{DG}_{t,i} \le P^{DG,max}_{i}.\;x^{DG}_{t,i}
\end{equation}
\begin{equation}\label{QDGlimit}
Q^{DG,min}_{i}.\;x^{DG}_{t,i} \le Q^{DG}_{t,i} \le Q^{DG,max}_{i}.\;x^{DG}_{t,i}
\end{equation}
\begin{equation}\label{DGSOCP}
(P^{DG}_{t,i})^2 + (Q^{DG}_{t,i})^2 \le (S^{DG})^2
\end{equation}
\begin{equation}\label{DGSU1}
C^{SU}_{t,i} \ge (x^{DG}_{t,i}-x^{DG}_{t-1,i}).\;\rho^{SU}
\end{equation}
\begin{equation}\label{DGSU2}
C^{SU}_{t,i} \ge 0
\end{equation}
\begin{equation}\label{DGSD1}
C^{SD}_{t,i} \ge (x^{DG}_{t-1,i}-x^{DG}_{t,i}).\;\rho^{SD}
\end{equation}
\begin{equation}\label{DGSD2}
C^{SD}_{t,i} \ge 0
\end{equation}
\begin{equation}\label{DGramp}
-R^D_i \le P^{DG}_{t,i} - P^{DG}_{t-1,i} \le R^U_i
\end{equation}
Considering the above-mentioned equations, $x^{DG}_{t,i}$, $P^{DG}_{i,t}$, $Q^{DG}_{t,i}$, $b^{DG}$, $k^{DG}$, and $C^{DG}_{t,i}$ denote status of DGs as binary variable, active/reactive power output of DGs, fixed and ramping cost of DGs, and operational cost of DG units, respectively. Additionally, $C^{SU}_{t,i}$, $C^{SD}_{t,i}$, $\rho^{SU}$, $\rho^{SD}$, $R^U_i$, and $R^D_i$ refer to startup/shutdown cost variables, and ramp up and ramp down limits, respectively. 

\subsubsection{Operational Constraints of PV Units}
The operational costs of utility-operated PV units, minimum and maximum limit of output power, and inverter capacity constraints are presented in equations (\ref{PVcost})-(\ref{PVreactive}), respectively. 

\begin{equation}\label{PVcost}
C^{PV}_{t,i} = c^{PV}.\;P^{PV}_{t,i}
\end{equation}
\begin{equation}\label{PVlimit}
0 \le P^{PV}_{t,i} \le P^{PV, max}
\end{equation}
\begin{equation}\label{PVreactive}
(P^{PV}_{t,i})^2 + (Q^{PV}_{t,i})^2 \le (S^{PV})^2
\end{equation}
where $C^{PV}_{t,i}$, $c^{PV}$, $P^{PV}_{t,i}$ refer to operational cost and output power of PV units, respectively. Additionally, $P^{PV, max}$, $Q^{PV}_{t,i}$, and $S^{PV}$ refer to maximum output power limitation, reactive power, and inverter size of PV units, respectively.  

\subsubsection{Operational Constraints of H2 Systems}
The operational constraints of H2 systems are presented in (\ref{ELX})-(\ref{HSinverter}). Let us denote $P^{EL}_{t,i}$, $P^{FC}_{t,i}$,  $QH^{EL}_{t,i}$, and $QH^{FC}_{t,i}$ as electrolyzer consumed power, FC generated power, amount of H2 consumed by electrolyzer in $kg$, and generated power by FC units in $MWh$, respectively. Additionally, $\pi^{EL}$, $\pi^{FC}$, $\eta^{EL}$, $\eta^{FC}$ refer to converting factors for electrolyzer and FC units, and efficiencies of these assets in the network. Equations (\ref{ELX})-(\ref{FClimit}) refer to the H2 production/consumption level of electrolyzer/stationary FC units based on efficiencies and converting factors. Additionally, these constraints prevent the simultaneous operation of electrolyzer and FC units by considering a binary variable $\psi^{HS}_{t,i}$. H2 mass balance, denoted as $MOH_{t,i}$, considering the transportation demand from FCEVs ($QH^{dem}_{t,i}$) and dissipation rate ($\pi^{Dsp}$), as well as storage tank capacity limits are expressed in (\ref{MOH}) and (\ref{MOHlimit}), respectively. Moreover, constraint (\ref{HSinverter}) expresses the H2 systems inverter for reactive power support, in which $Q^{HS}_{i,t}$ and $S^{HS}$ refer to reactive power of H2 systems and inverter size, respectively. Finally, electrolyzer capacity factor ($CF_i^{EL}$) during the optimization horizon and limits on electrolyzer load curtailment ($P_{t,i}^{EL,Shd}$) can be calculated based on (\ref{ELXCF}) and (\ref{shedlimit}), respectively. \par

\begin{equation}\label{ELX}
QH^{EL}_{t,i}= \pi^{EL}.\; P^{EL}_{t,i} .\;\eta^{EL}
\end{equation}
\begin{equation}\label{FC}
P^{FC}_{t,i}= \pi^{FC}.\; QH^{FC}_{t,i} .\;\eta^{FC}
\end{equation}
\begin{equation}\label{ELXlimit}
QH^{EL,min}_{i}.\;\psi^{HS}_{t,i}\le QH^{EL}_{t,i} \le QH^{EL,max}_{i}.\;\psi^{HS}_{t,i}
\end{equation}
\begin{equation}\label{FClimit}
QH^{FC,min}_{i}.\;(1-\psi^{HS}_{t,i}) \le QH^{FC}_{t,i} \le QH^{FC,max}_{i}.\;(1-\psi^{HS}_{t,i})
\end{equation}
\begin{equation}\label{MOH}
\begin{split}
MOH^{H2}_{t,i}= & MOH^{H2}_{t-1,i}\;-\pi^{Dsp}.\; MOH^{H2}_{t,i}\;+ (QH^{EL}_{t,i} \\ &- QH^{dem}_{t,i} - QH^{FC}_{t,i} ).\;\Delta t
\end{split}
\end{equation}
\begin{equation}\label{MOHlimit}
MOH^{H2, min}_{i}\le MOH^{H2}_{t,i} \le MOH^{H2, max}_{i}
\end{equation}
\begin{equation}\label{HSinverter}
(P^{EL}_{i,t}-P^{FC}_{i,t})^2 + (Q^{HS}_{i,t})^2 \le (S^{HS})^2
\end{equation}
\begin{equation}\label{ELXCF}
CF^{EL}_{i} = \frac{ \sum_{t=1}^{T} P^{EL}_{t,i}}{\sum_{t=1}^{T} P^{EL,max}_{i}}
\end{equation}
\begin{equation}\label{shedlimit}
0 \le P^{EL,Shd}_{t,i} \le \frac{QH^{dem}_{t,i}}{\pi^{EL}.\;\eta^{EL}}
\end{equation}

To show the flexibility of H2 systems during the normal and emergency operations, three CBDRS are considered. In (\ref{DRELX}) and (\ref{DRFC}), CBDR signals, shown as $PDR^{Sgl}$, are expressed for electrolyzers and FC units. In the case of any external signal from DSO, based on their available capacity (electrolyzer, storage tank, and FC units), H2 systems follow the signal and act as load or generation asset based on equations (\ref{DRELX}) and (\ref{DRFC}). Accordingly, in the case of $N-m$ contingencies, H2 systems can act as long-term energy storage with long-duration times compared to batteries. To that end, constraint (\ref{DRMOH}) expresses the demand response (DR) signal, in which $\kappa_t$ denotes the percentage of H2 required from DSO regarding the H2 mass in the tank, as a reserve before emergency operation. Prior to any forecasted disruption ($t < t_{event}$), DSO asks H2 system owners to fill their tank completely as a backup generation unit for supplying the load in the post-event time ($t \ge t_{event}$). This will help DSO to minimize the total cost and total load curtailment during $N-m$ contingencies.  
\begin{equation}\label{DRELX}
\begin{split}
 Sgn(PDR^{Sgl})\;.\;PDR^{Sgl} \le \sum_{i=1}^{N_H}P^{EL}_{t,i} \le & \sum_{i=1}^{N_H}P^{EL,max},\;\; \\ & if \; PDR^{Sgl} \ge 0.
\end{split}
\end{equation}
\begin{equation}\label{DRFC}
\begin{split}
 Sgn(PDR^{Sgl})\;.\;PDR^{Sgl} \le \sum_{i=1}^{N_H}P^{FC}_{t,i} \le & \sum_{i=1}^{N_H}P^{FC,max},\;\; \\ & if \; PDR^{Sgl} \le 0.
 \end{split}
\end{equation}
\begin{equation}\label{DRMOH}
 \sum_{i=1}^{N_H} MOH^{H2}_{t,i} \ge \kappa_{t}\;.\;\sum_{i=1}^{N_H} MOH^{H2,max}
\end{equation}

\subsubsection{SOCP-based Distribution Network AC Power Flow Model}
The AC power flow constraints (addressing both normal and emergency operation) based on branch flow model are presented in (\ref{voltage})-(\ref{shedreactive}). The voltage constraints are shown in (\ref{voltage}) and (\ref{voltagelimit}), in which $V_{t,i}$ and $a_{t,l}$ represent voltage values for each node and current values for each branch, respectively. $R$ and $X$ are the resistance and reactance of lines. Let us denote $f_{t,i}^{p/q}$, $P_{t,i}^{load}$, $Q_{t,i}^{load}$, $P_{t,i}^{Load,shd}$, $Q_{t,i}^{Load,shd}$ as the active/reactive power flow of lines, active/reactive load of DN, and active/reactive amount of load curtailment, respectively. Additionally, $P_{t,i}^{Load,shd}$ denotes the amount of curtailed load by electrolyzers (equivalent FCEVs demand in MW). The active and reactive power balance equations are shown in (\ref{activebalance}) and (\ref{reactivebalance}), respectively. Line flows are limited by equations (\ref{linelimit}) and (\ref{linelimit1}), and SOCP-based constraints are presented in (\ref{SOCP}). More details regarding the exact conic relaxation can be found in \cite{farivar2013branch}. Finally, equations (\ref{shedcost})-(\ref{shedreactive}) express the constraints for emergency operation which may result in load curtailment. It should be noted that load curtailment is penalized by the value of loss of load ($VOLL$) , based on the load importance. For instance, this value is \$10,000/MWh for critical loads , \$5,000/MWh for moderately-critical loads, and \$1,000/MWh for non-critical loads. 
\begin{equation}\label{voltage}
V_{t,i} = V_{t,j} - 2(R_{ji}\;.\;f^p_{i,t} -X_{ji}\;.\;f^q_{t,i}) + (R_{ji}^2 + X_{ji}^2)\;.\; a_{t,l}
\end{equation}
\begin{equation}\label{voltagelimit}
(V^{min})^2 \le V_{t,i} \le (V^{max})^2
\end{equation}
\begin{equation}\label{activebalance}
\begin{split}
f^p_{t,i} & =  P^{Load}_{t,i} + \sum_{j\rightarrow i} f^p_{t,j} + R_{ji}.a_{t,l} + P^{EL}_{t,i} + P_{t,i}^{Exb} - P_{t,i}^{Exs}  \\ & - P^{FC}_{t,i} - P^{PV}_{t,i} - P^{DG}_{t,i} - P^{EL,Shd}_{t,i} - P^{Load,Shd}_{t,i}    
\end{split}
\end{equation}
\begin{equation}\label{reactivebalance}
\begin{split}
f^q_{t,i} & = Q^{Load}_{t,i} + \sum_{j\rightarrow i} f^q_{t,j} + X_{ji}.a_{t,l} + Q^{EL}_{t,i} - Q^{FC}_{t,i} - Q^{PV}_{t,i} \\ & - Q^{DG}_{t,i} - Q^{Load,Shd}_{t,i} + Q^{HS}_{t,i}     
\end{split}
\end{equation}
\begin{equation}\label{linelimit}
(f^p_{t,i})^2 + (f^q_{t,i})^2 \le (S^{line})^2
\end{equation}
\begin{equation}\label{linelimit1}
(f^p_{t,i} -  R_{ji}\;.\; a_{t,l} )^2 + (f^q_{t,i}-  X_{ji}\;.\; a_{t,l} )^2 \le (S^{line})^2
\end{equation}
\begin{equation}\label{SOCP}
[(f^p_{t,i})^2 + (f^q_{t,i})^2]\;.\;\frac{1}{a_{t,l}} \le V_{t,i}
\end{equation}
\begin{equation}\label{shedcost}
C^{Load,Shd}_{t,i} = VOLL(i) .\;P^{Load,Shd}_{t,i}
\end{equation}
\begin{equation}\label{shedlimit}
0 \le P^{Load,Shd}_{t,i} \le P^{Load}_{t,i}
\end{equation}
\begin{equation}\label{shedreactive}
Q^{Load,Shd}_{t,i} = P^{Load,Shd}_{t,i}\;.\;\frac{Q^{Load}_{t,i}}{P^{Load}_{t,i}}
\end{equation}

\subsection{Lower-Level Primal Problem: TSO Perspective}
The objective function and constraints of LL primal problem are presented in (\ref{TSOOF})-(\ref{exchangebuy}). The objective of TSO is to maximize the social welfare, or equivalently minimizing the operation cost. In (\ref{TSOOF}), $Pg_{t,i}$, $\rho^{b/s}$, and $Pw_{t,i}$ are active power of generation units, the offered and bid price from DSO, dispatched wind power, respectively. In order to integrate the dual of LL primal problem into UL problem, dual variables are assigned to all equations (\ref{Tpowerbalance})-(\ref{exchangebuy}). Equations (\ref{Tgenlimit}) and (\ref{Tpowerbalance}) show the generator minimum and maximum generation limits, and power balance for transmission network, respectively. It should be noted that $TD_{t,i}$ refers to TN load. Additionally, (\ref{Tlineflow}) and (\ref{Tlinelimit}) model the line flow denoted as $Tfl_{t,i}$ and its thermal limits based on DC power flow, in which $\delta$ denote the voltage phase angle. Moreover, wind power constraint is presented in (\ref{windspill}). Finally, the active power exchange between DN and TN are constrained by (\ref{exchangesell}) and (\ref{exchangebuy}).
\begin{equation}\label{TSOOF}
\begin{split}
\text{min.}\; \sum_{t=1}^{T} & \Bigg\{\sum_{i=1}^{N_G} C_i^g Pg_{t,i} - \rho_{t}^{b}\;P_{t,i}^{Exb}+ \rho_{t}^{s}\;P_{t,i}^{Exs} \\&  + \sum_{i=1}^{N_W} C_i^w \; Pw_{t,i} \Bigg\}
\end{split}
\end{equation}
\begin{equation}\label{Tgenlimit}
Pg^{min}_i\le Pg_{t,i} \le Pg^{max}_i\;\;:(\underline{\alpha}_{t,i},\overline{\alpha}_{t,i})
\end{equation}
\begin{equation}\label{Tpowerbalance}
\begin{split}
\sum_{i=1}^{NG}Pg_{t,i} & + \sum_{i\rightarrow j} Tfl^p_{t,i} - \sum_{j\rightarrow i} Tfl^p_{j,t} - P_{t,i}^{Exb} + P_{t,i}^{Exs} + Pw_{t,i} \\& = TD_{t,i}\;\;: (\lambda_{t,i})
\end{split}
\end{equation}
\begin{equation}\label{Tlineflow}
Tfl_{t,i} = \frac{1}{X_l}(\delta_i-\delta_j)\;\;:(\zeta_{t,i})
\end{equation}
\begin{equation}\label{Tlinelimit}
TFl^{min}_l\le Tfl_{t,l} \le TFl^{max}_l\;\;:(\underline{\delta}_{t,l},\overline{\delta}_{t,l})
\end{equation}
\begin{equation}\label{windspill}
Pw_{t,i} \le Pw_{t,i}^{max}\;\;: (\gamma_{i,t})
\end{equation}
\begin{equation}\label{exchangesell}
P_{t,i}^{Exs}\le P^{UG,max} \;.\;U_{t,i}\;\;: (\overline{\psi}_{t,i})
\end{equation}
\begin{equation}\label{exchangebuy}
P_{t,i}^{Exb} \le P^{UG,max}\;.\; (1-U_{t,i})\;\;: (\underline{\psi}_{t,i})
\end{equation}

\subsection{Lower-Level Dual Problem: TSO Perspective}
The duality-based technique \cite{arroyo2010bilevel} is used to solve the aforementioned bi-level problem, by integrating the dual of LL problem into UL problem and achieving a single-level equivalent problem. The following equations are the dual problem of equations (\ref{TSOOF})-(\ref{exchangebuy}).
\begin{equation}\label{dualLLOF}
\begin{split}
\text{max.}\;& \sum_{t=1}^{T} \Bigg\{ \sum_{i=1}^{N_G}(Pg^{min}_i\underline{\alpha}_{t,i} + Pg^{max}_i\overline{\alpha}_{t,i}) + \sum_{l=1}^{N_l}(TFl^{min}_l\underline{\delta}_{t,l} \\ &  + TFl^{max}_l \overline{\delta}_{t,l}) \sum_{i=1}^{N_i}( P^{UG,max}\;(1-U_{t,i})\;\underline{\psi}_{t,i} \\ &  +  P^{UG,max}\;U_{t,i}\;\overline{\psi}_{t,i}) + \sum_{i=1}^{N_i}\lambda_{t,i}TD_{t,i} + \sum_{i=1}^{N_i}\gamma_{i,t}Pw_{t,i} \Bigg \} 
\end{split}
\end{equation}
\begin{equation}\label{dualLL1}
\overline{\alpha}_{t,i} + \underline{\alpha}_{t,i} + \lambda_{i,t} = C_i^g
\end{equation}
\begin{equation}\label{dualLL2}
\overline{\delta}_{t,l} + \underline{\delta}_{t,l} + \zeta_{t,l} + \lambda_{t,i} -\lambda_{t,j} = 0
\end{equation}
\begin{equation}\label{dualLL3}
\overline{\psi}_{t,i}+\lambda_{t,i} \le \rho_{t}^{s}
\end{equation}
\begin{equation}\label{dualLL4}
\underline{\psi}_{t,i}-\lambda_{t,i} \le \rho_{t}^{b}
\end{equation}
\begin{equation}\label{dualLL5}
-\sum_{l=ord(i)}^{N_L}\frac{\zeta_{t,l}}{X_l} + \sum_{l=ord(j)}^{N_L}\frac{\zeta_{t,l}}{X_l} =0
\end{equation}
\begin{equation}\label{dualLL6}
\gamma_{t,i} + \lambda_{t,i} \le C_i^w 
\end{equation}

\subsection{Strong Duality for Primal and Dual Problems of TSO}
The strong duality theory is applied on primal and dual LL problems to obtain the optimal solution, as expressed in equation (\ref{strongduality}).
\begin{equation}\label{strongduality}
\begin{split}
& \sum_{t=1}^{T}\Bigg\{\sum_{i=1}^{N_G} C_i^g Pg_{t,i} - \rho_{t}^{b}P_{t,i}^{Exb}+ \rho_{t}^{s} P_{t,i}^{Exs}+ \sum_{i=1}^{N_W} C_i^w\; w_{t,i} \Bigg\} \\ &= \sum_{t=1}^{T} \Bigg\{ \sum_{i=1}^{N_G}(Pg^{min}\underline{\alpha}_{t,i} + Pg^{max}\overline{\alpha}_{t,i}) + \sum_{l=1}^{N_l}(TFl^{min}\underline{\delta}_{t,l} \\ & + TFl^{max} \overline{\delta}_{t,l}) +   \sum_{i=1}^{N_i}( P^{UG,max}\;(1-U_{t,i})\;\underline{\psi}_{t,i} \\ & +  P^{UG,max}\;U_{t,i}\;\overline{\psi}_{t,i}) + \sum_{i=1}^{N_i}\lambda_{t,i}TD_{t,i}  \sum_{i=1}^{N_i} + \gamma_{t,i} Pw_{t,i} \Bigg \}  
\end{split}
\end{equation}

\subsection{Single-Level Equivalent Optimization}
After integrating the dual of LL problem into the UL problem, the single-level equivalent problem is as follows:
\begin{equation}\label{dualLLOF}
\begin{split}
\text{min.}\; \text{Equation (\ref{ULOF})} \qquad \qquad \qquad \qquad \qquad \quad\\
 \text{s.t} \qquad \qquad \qquad \qquad \qquad \qquad \qquad  \\
 \text{Equations (\ref{DGcost})-(\ref{shedreactive}), (\ref{Tgenlimit})-(\ref{exchangebuy}), (\ref{dualLL1})-(\ref{strongduality})} 
\end{split}
\end{equation}

\subsection{H2 Production Cost Calculation for Optimization Horizon}
Based on the outcome of the aforementioned optimization problem and the solution technique presented in \cite{papavasiliou2017analysis}, DLMP for each node including power loss, voltage regulation, and congestion costs of DN can be calculated for each node of DN. As a result, H2 production cost including water electrolysis and storage cost can be calculated based on Algorithm 1.

\begin{figure}
\centering
\footnotesize
\captionsetup{justification=raggedright,singlelinecheck=false,font={footnotesize}}
	\includegraphics[width=3.5in,height=2.4in]{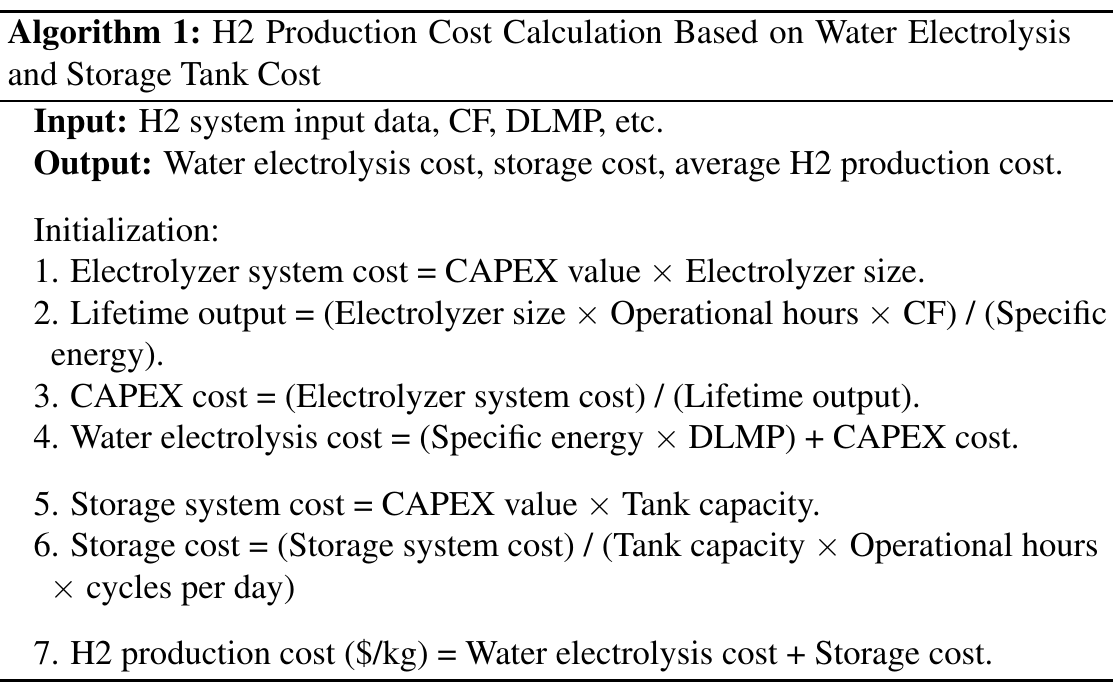}
    \label{algorithm}
\end{figure}

\section{Simulation Results and Analysis}\label{results}
\begin{figure}
\centering
\footnotesize
\captionsetup{justification=raggedright,singlelinecheck=false,font={footnotesize}}
	\includegraphics[width=3.5in, height=2in]{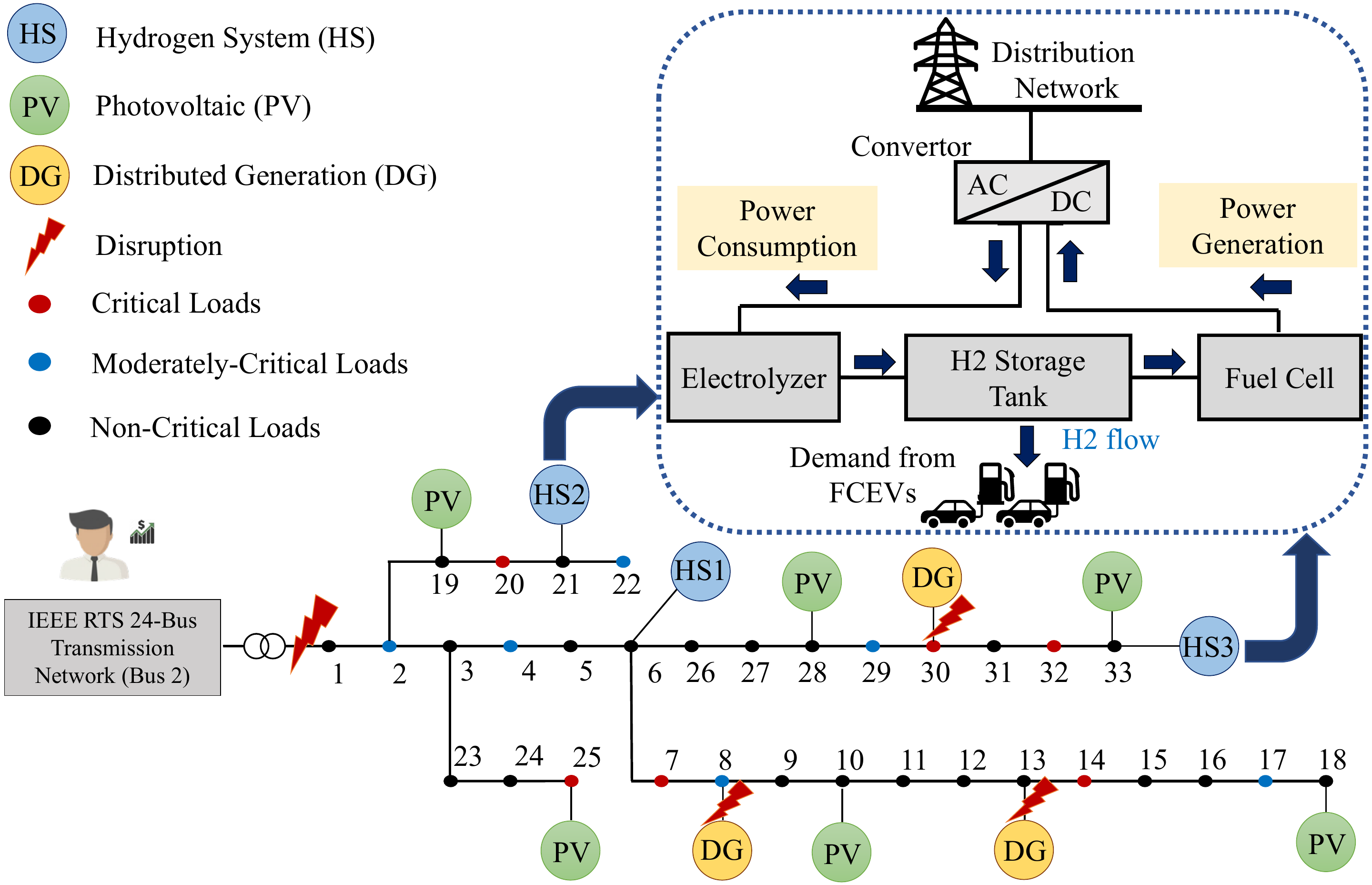}
	\caption{33-node distribution test system which is connected to bus 2 of IEEE RTS 24-bus test system.}
    \label{testsystem}
\end{figure}

\begin{figure}
\centering
\footnotesize
\captionsetup{justification=raggedright,singlelinecheck=false,font={footnotesize}}
	\includegraphics[width=3.5in,height=1.4in]{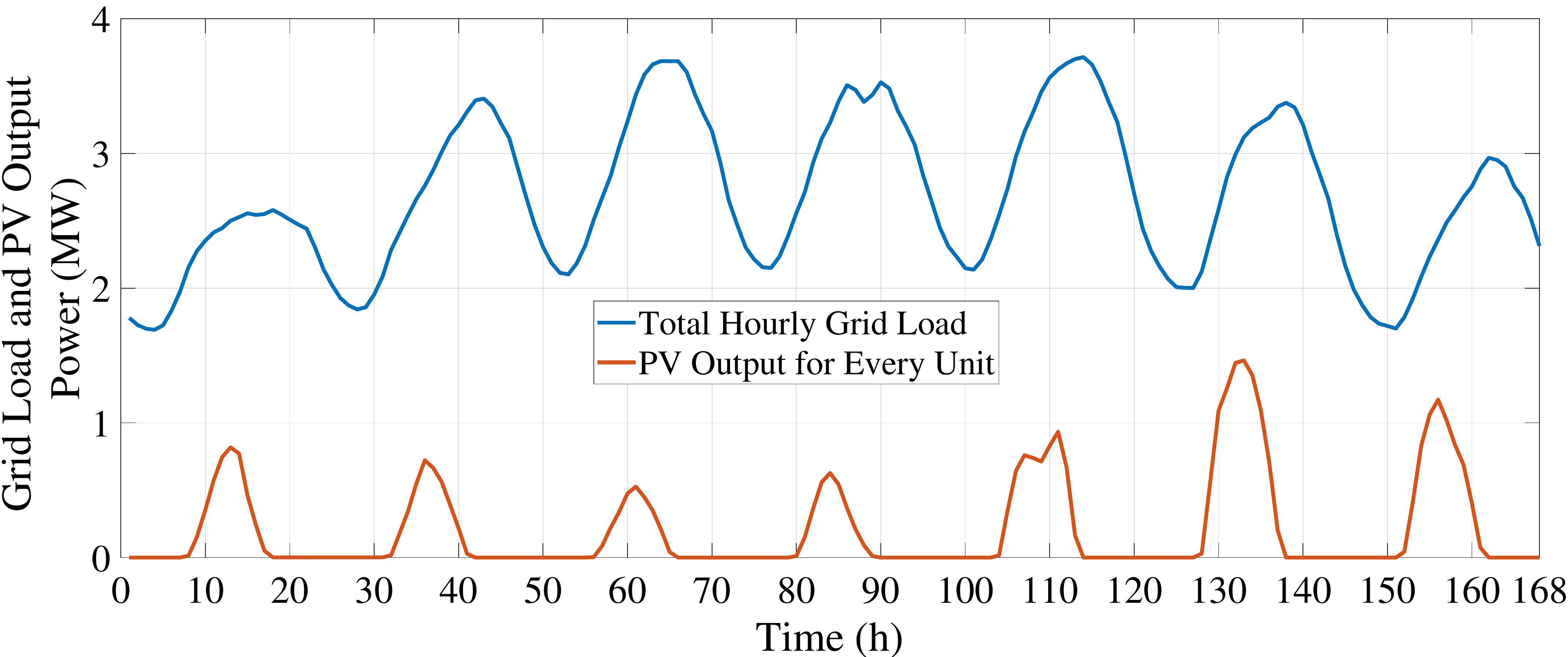}
		\caption{Hourly 33-node load pattern and output PV power pattern.}
    \label{loadprofile}
\end{figure}

\begin{figure}
\centering
\footnotesize
\captionsetup{justification=raggedright,singlelinecheck=false,font={footnotesize}}
	\includegraphics[width=3.5in]{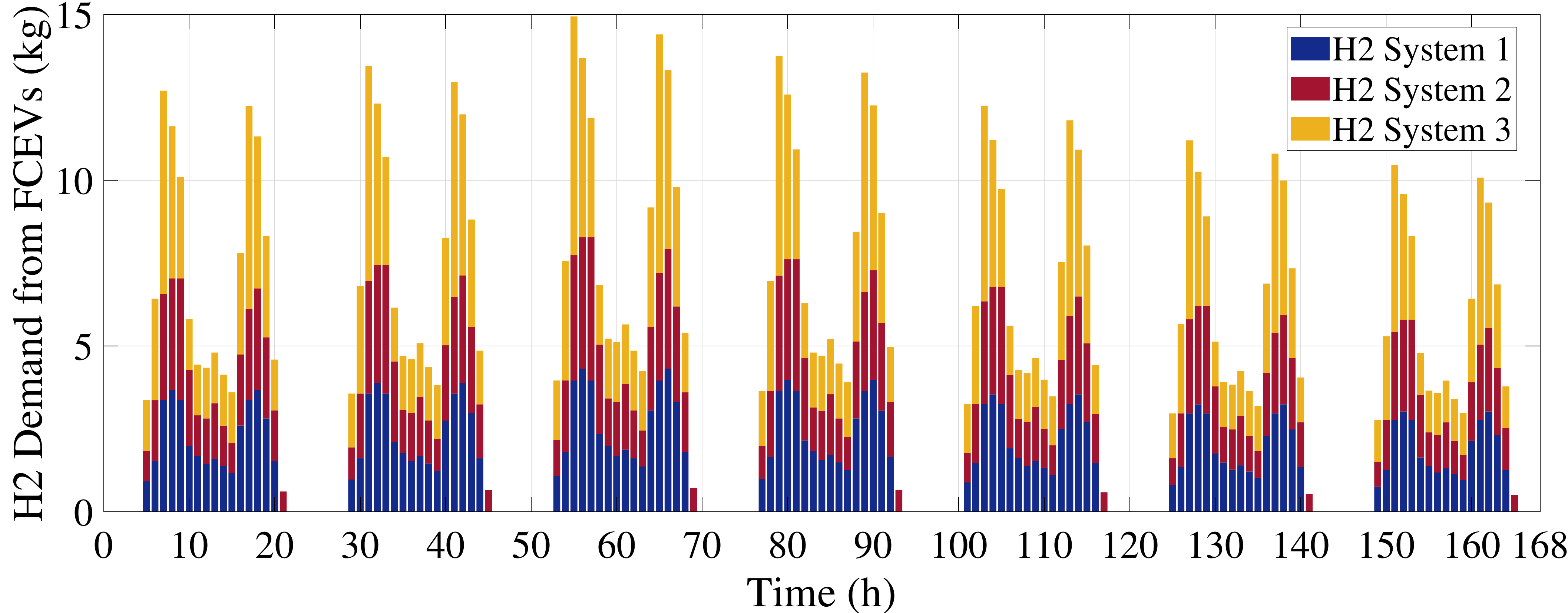}
		\caption{Hourly hydrogen demand from transportation sector.}
    \label{HDE}
\end{figure}

The proposed method is validated by testing on IEEE 24-bus transmission test system \cite{ordoudis2016updated}, and 33-node distribution test feeder \cite{baran1989network} with an hourly time step during the week. The transmission network hosts six wind farms located at buses 3, 5, 7, 16, 21, and 23. More information regarding the transmission system and wind farms' capacity can be found in \cite{ordoudis2016updated}. The distribution network includes three natural gas power plants, six utility-operated PV units with the total capacity same as total grid load (based on scaling factors), and three H2 systems, as shown in Fig. \ref{testsystem}. The operational costs including capital expenditures (CAPEX) are considered for generation assets based on National Renewable Energy Laboratory's (NREL) advanced technology baseline \cite{vimmerstedt20182018}. The H2 demand requested by FCEVs is calculated based on the method presented in \cite{sun2018optimal}. These FCEVs are considered as Honda Clarity models \cite {FCEV} assuming that these cars arrive at the H2 fueling station with 45\% H2 fuel in their tanks. Moreover, the load, solar (without scaling factor), and FCEVs weekly patterns are shown in Fig. \ref{loadprofile} and Fig. \ref{HDE}, respectively. Different from other research efforts focused on line outages, in order to show the benefits of H2 systems over batteries, it is assumed that the tie line connecting distribution network to transmission network as well as three natural gas power plants are out of service for almost two days. The rolling optimization horizon is 48 hours, in which the DSO release the day-ahead operation based on the next 24 hours data forecasts. The electrolyzer and FC units sizes are considered as 0.5 MW. Additionally, the storage tank capacity, specific energy, electrolyzer and FC efficiencies are from \cite{haggi2021proactive}.\par

The simulations are carried out on a PC with an Intel Core-i7 CPU of 1.8 GHz and 16 GB RAM. The proposed framework is solved using GAMS/Gurobi \cite{rosenthal2016gams} with a gap of 0.1\%.\par

\begin{figure}
\centering
\footnotesize
\captionsetup{justification=raggedright,singlelinecheck=false,font={footnotesize}}
	\includegraphics[width=3.5in]{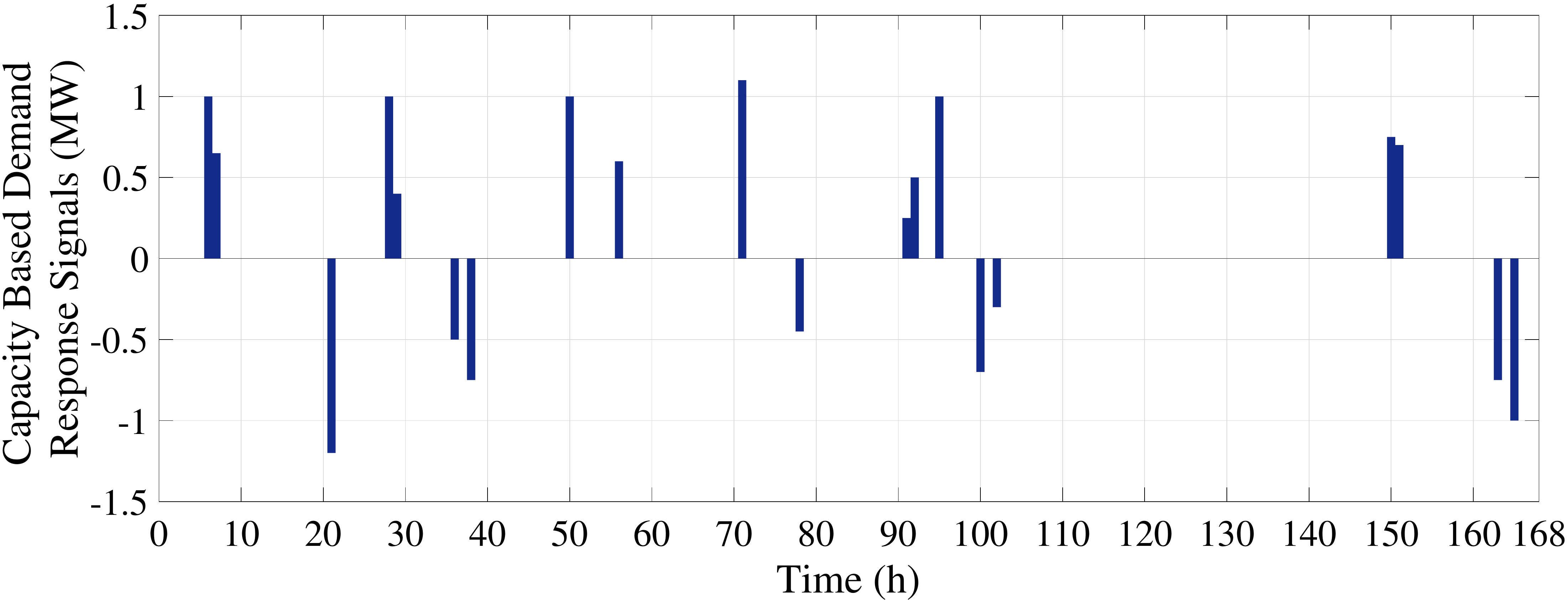}
		\caption{Demand response signals imposed to H2 systems by DSO. }
    \label{CBDR}
\end{figure}

\vspace{-0.2cm}
\subsection{Results for Operation of Integrated Distribution and Transmission Networks}
\begin{figure}
\centering
\footnotesize
\captionsetup{justification=raggedright,singlelinecheck=false,font={footnotesize}}
	\includegraphics[width=3.5in]{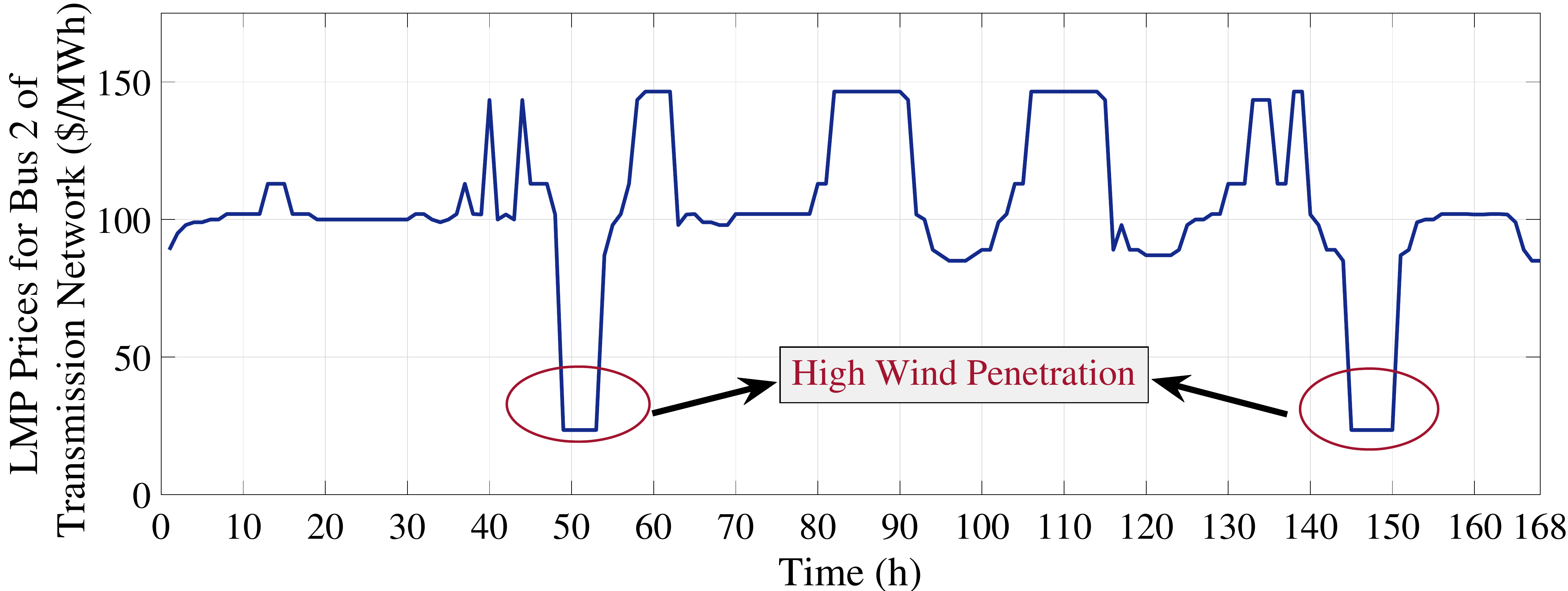}
		\caption{LMP prices,$\lambda$, from transmission network. }	
    \label{LMP}
\end{figure}

Considering the coordinated operation of distribution and transmission systems, Fig. \ref{LMP} shows the hourly LMP prices of bus 2 of transmission systems, which TSO shares with DSO for power exchanges. As it can be seen, due to the different participation levels of generation units located in transmission network, different LMP values are obtained. Additionally, due to the high wind penetration during hours 49 to 53 and 145 to 150, LMP values are \$23.5/MWh. In these hours, TSO sells energy to distribution system since the lowest DG operational cost is \$36/MWh. Moreover, in the case of emergency, DSO must purchase energy from TSO in the case that total DN generation capacity is not sufficient. More details will be presented in the following section.

\subsection{Results for Proactive Scheduling of H2 Systems with and without Rolling Horizon Approach}
\begin{figure}
\centering
\footnotesize
\captionsetup{justification=raggedright,singlelinecheck=false,font={footnotesize}}
	\includegraphics[width=3.5in]{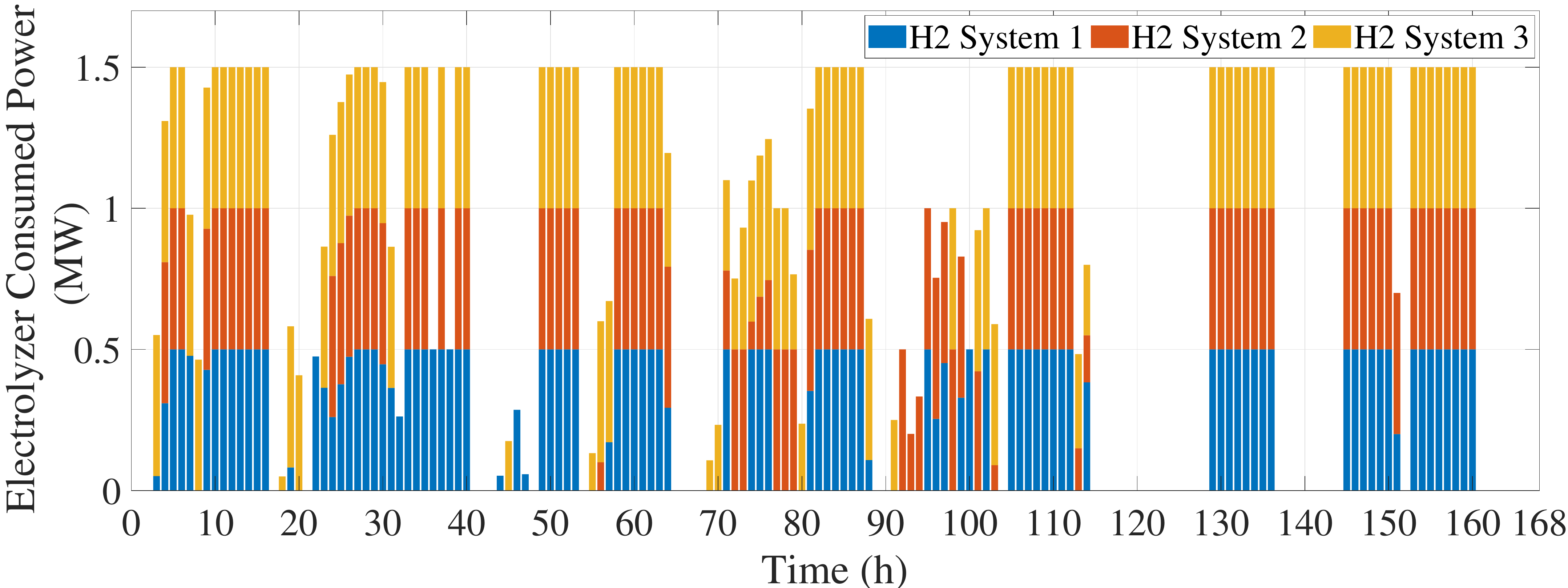}
		\caption{Electrolyzer scheduling w/o considering rolling horizon approach.}
    \label{Weekly_ELX_NoRolling}
\end{figure}

\begin{figure}
\centering
\footnotesize
\captionsetup{justification=raggedright,singlelinecheck=false,font={footnotesize}}
	\includegraphics[width=3.5in]{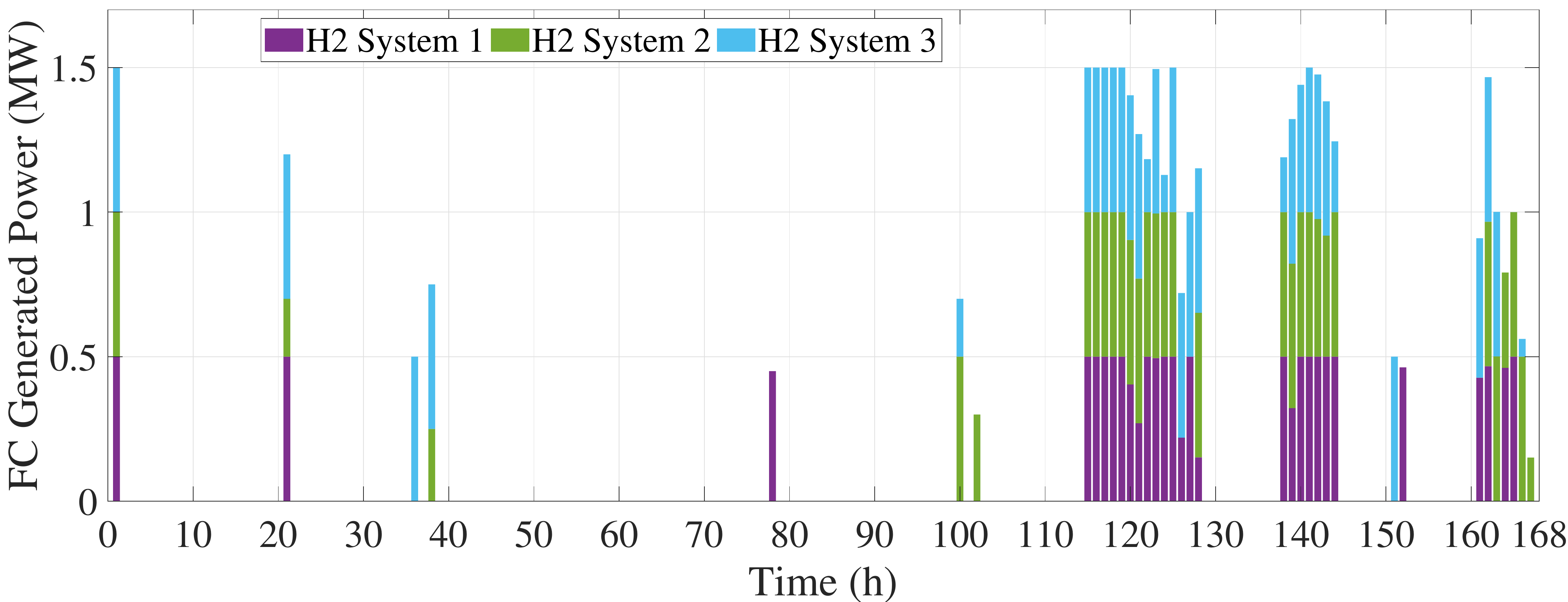}
		\caption{FC units scheduling w/o considering rolling horizon approach.}		
    \label{Weekly_FC_NoRolling}
\end{figure}

\begin{figure}
\centering
\footnotesize
\captionsetup{justification=raggedright,singlelinecheck=false,font={footnotesize}}
	\includegraphics[width=3.5in]{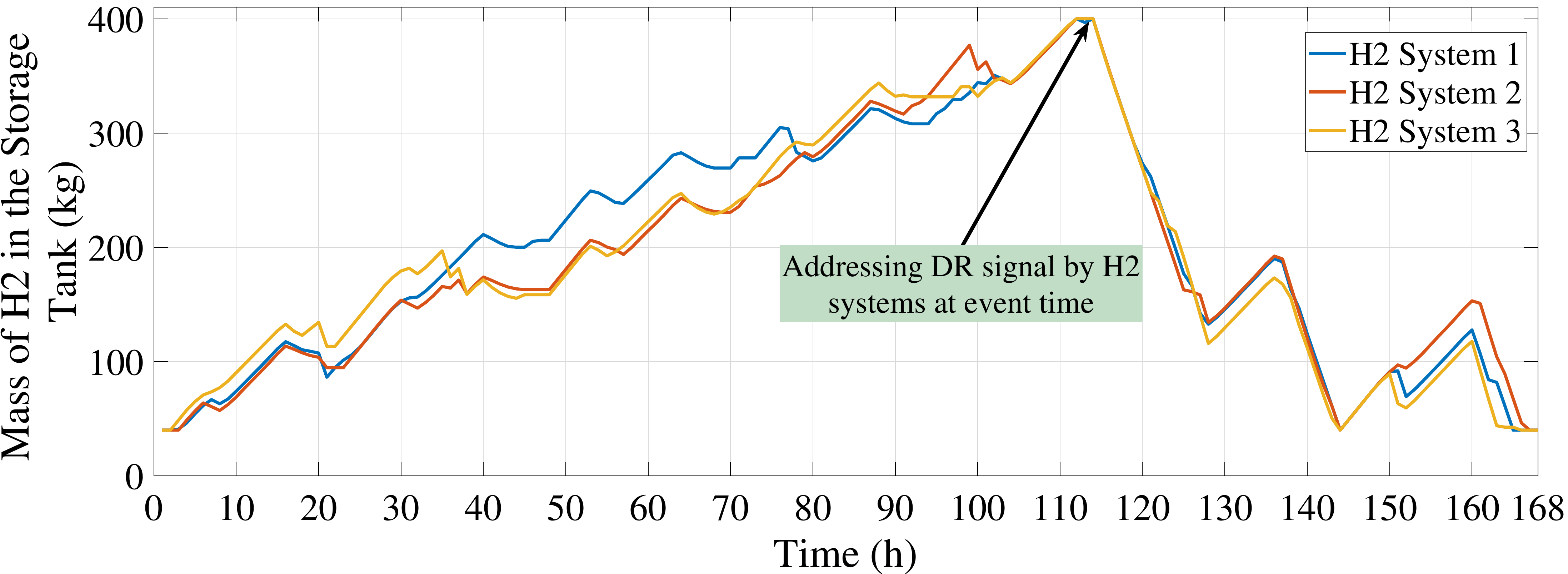}
		\caption{Mass of H2 in storage tank w/o considering rolling horizon approach.}
    \label{Weekly_LOH_NoRolling}
\end{figure}

\begin{figure}
\centering
\footnotesize
\captionsetup{justification=raggedright,singlelinecheck=false,font={footnotesize}}
	\includegraphics[width=3.5in]{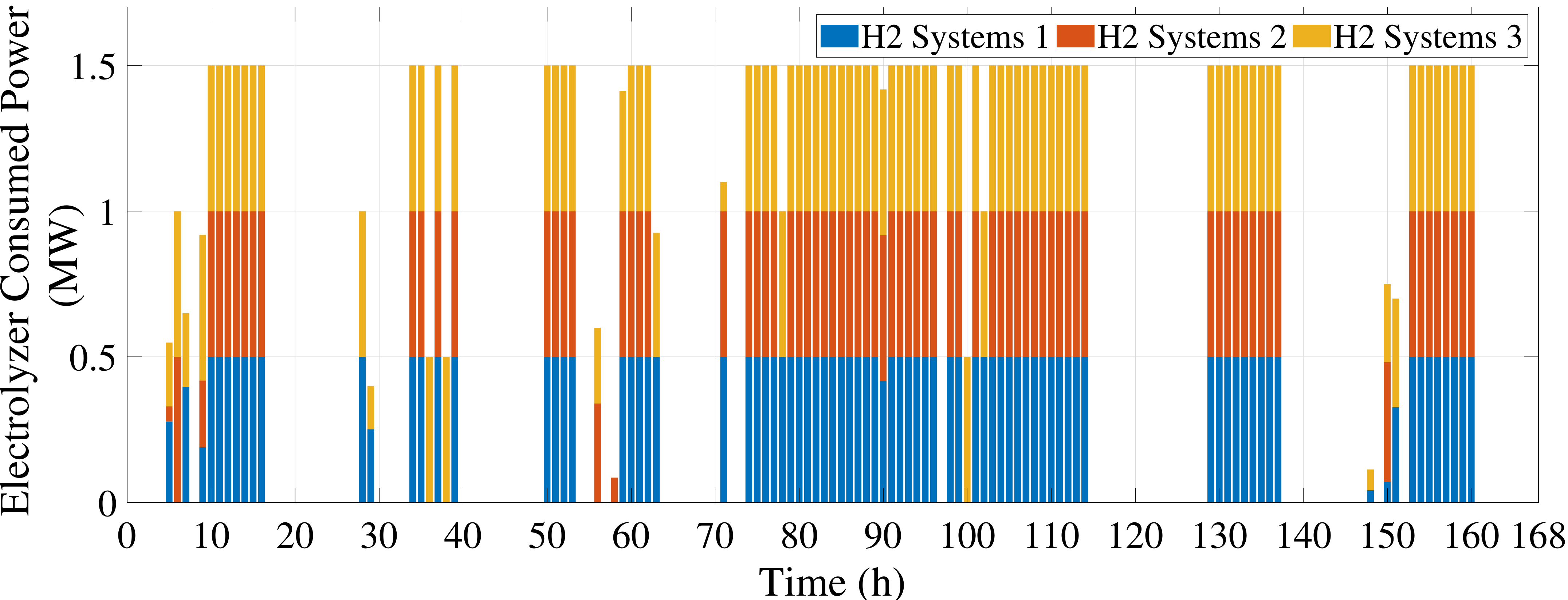}
		\caption{Electrolyzer scheduling considering rolling horizon approach.}
    \label{ELX_rolling}
\end{figure}

\begin{figure}
\centering
\footnotesize
\captionsetup{justification=raggedright,singlelinecheck=false,font={footnotesize}}
	\includegraphics[width=3.5in]{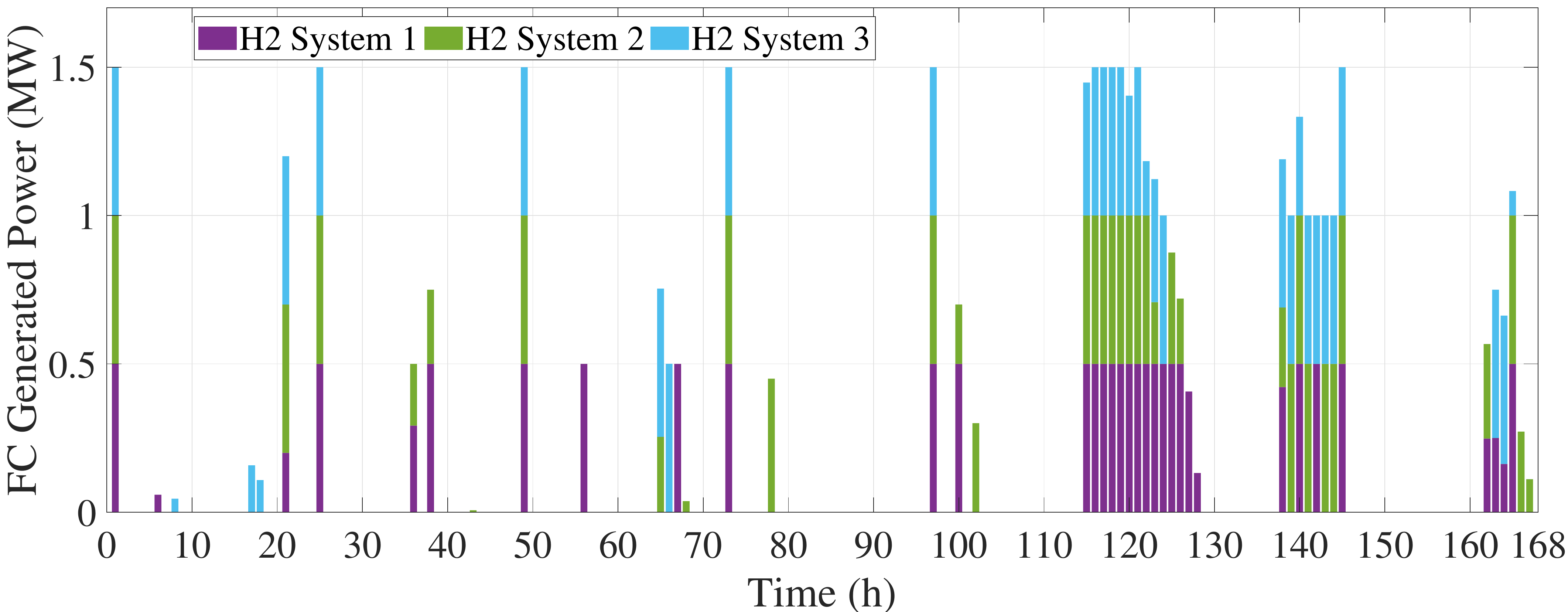}
		\caption{FC units scheduling considering rolling horizon approach.}
    \label{FC_rolling}
\end{figure}

\begin{figure}
\centering
\footnotesize
\captionsetup{justification=raggedright,singlelinecheck=false,font={footnotesize}}
	\includegraphics[width=3.5in]{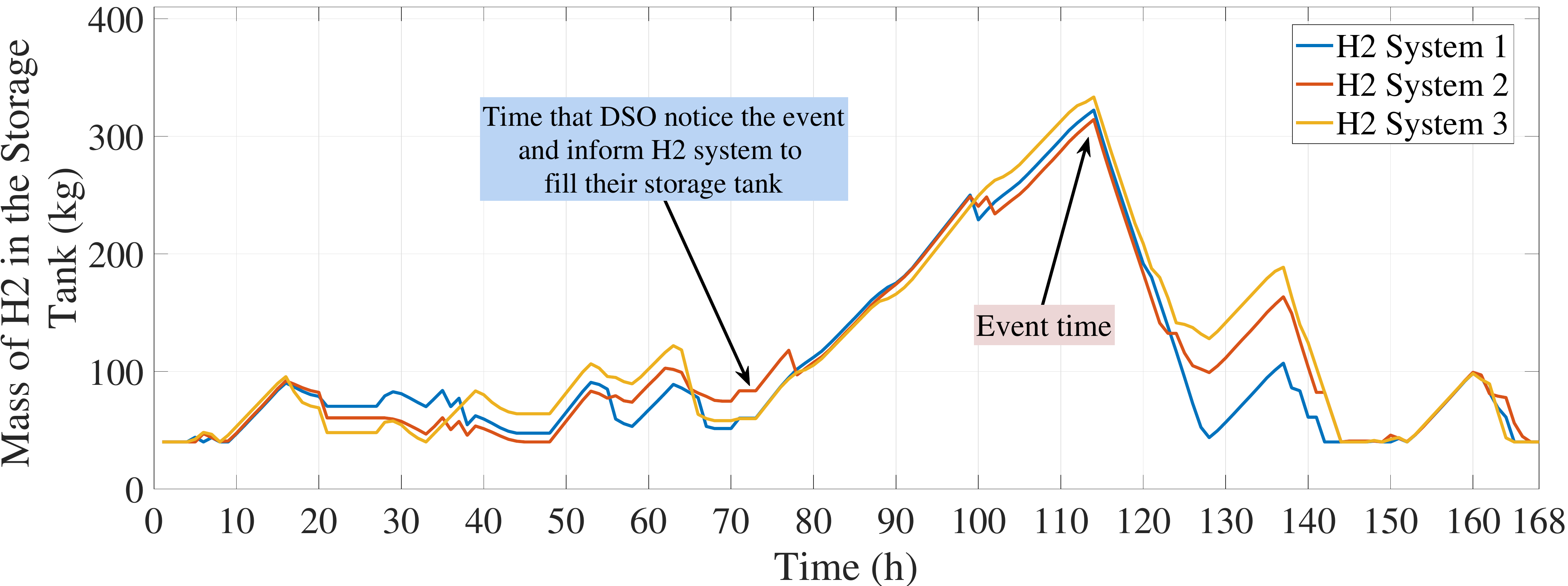}
		\caption{Mass of H2 in the storage tank considering rolling horizon approach.}
    \label{LOH_rolling}
\end{figure}

The results of proactive management of H2 systems with and without considering rolling horizon approach are shown in Fig. \ref{Weekly_ELX_NoRolling} to Fig. \ref{Weekly_LOH_NoRolling}, and Fig. \ref{ELX_rolling} to Fig. \ref{LOH_rolling}, respectively. In the proposed framework, H2 systems should follow the CBDR signals in both normal and proactive mode imposed by DSO, as depicted in Fig. \ref{CBDR}. It should be mentioned that positive and negative signals are for electrolyzers and FC units, respectively. In both scenarios (with and without rolling horizon), H2 systems respond to these signals. For instance, considering hours 21, 36 and 38, DSO asks the H2 systems to generate power for some purposes. Based on Fig. \ref{Weekly_FC_NoRolling} and Fig. \ref{FC_rolling}, it can be easily seen that these signals are addressed by H2 systems. It should be noted that the level of participation in CBDR signal is based on the technical reasons, such as the available capacity of electrolyzer, storage, FC, or economic reasons. On the other hand, DSO also sends signals to H2 systems to produce H2 (act as load in the system) due to the arrival of heavy duty H2 trucks, which requires at least 50kg of H2 for filling the tank \cite{H2truck}. For example, considering DR signal in hours 28 and 29, Fig. \ref{Weekly_ELX_NoRolling} and Fig. \ref{ELX_rolling} clearly show that these signals are addressed. \par

In the case of disruption, DSO must schedule its resources in advance to minimize the cost and load curtailment. However, sometimes the forecasts are inaccurate and proactive scheduling cannot reduce but even increase the operational cost, due to the increasing amount of reserve capacity. For instance, the hurricane direction is forecasted to hit the location, but changes the direction one or two days later. Fig. \ref{Weekly_LOH_NoRolling} shows the mass of H2 in the tank without considering rolling horizon. In this scenario, DSO has the access to perfect forecasts regarding the output power of PV units, the exact time of extreme event, etc., and imposes CBDR signals to H2 systems to fill their storage tank and be prepared for post-event times. That's the reason why the H2 mass in the tank gradually increases from hour 1 until hour 115 (when the extreme event happens). This can also be seen from Fig. \ref{Weekly_ELX_NoRolling}, in which from the first day of week, electrolyzers consume power to minimize the cost and address the CBDR signal regarding the extreme event. However, the aforementioned scenario is not applicable in real-world applications due to the reasons that DSO never has access to the perfect forecasts, and the expectations regarding the extreme event may not be true. To this end, Fig. \ref{LOH_rolling} shows the mass of H2 in the tank considering rolling horizon approach in which limits the access of DSO to the perfect forecasts regarding the extreme time and input data. As it can be seen, for the first 3 days, DSO normally supplies the grid load and transportation sector demand. However, in the rolling period of fourth day (which starts from hour 73 to 120), DSO sends the CBDR signal to H2 systems to fill their storage tank prior to hour 115 based on the available capacity of H2 system components. Different from the previous scenario, that's the reason why the H2 mass is not gradually increasing in the first 3 days. After the notice of DSO, H2 systems consume power to fill their tank as much as possible. This can also be seen in Fig. \ref{ELX_rolling} in which electrolyzers are fully operated from hours 73 to 114. Additionally, based on Fig. \ref{FC_rolling}, from hour 115, FC units start injecting power into the grid to supply the critical and moderately-critical loads. More details regarding the energy not supplied and the resilience index of H2 systems will be presented in the following sections.  


\subsection{Resilience Analysis for H2 systems Compared to Batteries with Different Duration Times}
To provide a fair comparison of the performance between H2 system and battery energy storage systems with different duration times, it is assumed that the transit sector demand is zero kg, and battery energy storage power ratings are the same as electrolyzers and FC units. Moreover, the battery efficiencies are considered as 90\%, and the constraints related to the CBDR signals are ignored. Five scenarios are considered in which the first four scenarios are focused on battery energy storage systems with 2, 4, 6, and 8 hours duration. The last scenario is for resilient scheduling with H2 systems. The total load of system during hours 115 till 144 is 84.3 MW, including 28.1 MW critical loads, 11.1 MW moderately-critical loads, and 45.1 MW non-critical loads. Table \ref{RsilienceTable} shows the energy not supplied for different load types for the aforementioned scenarios. This table clearly shows that using H2 systems prevents more load curtailment, especially for critical and moderately critical loads. Based on the results, H2 systems supplied 100\% of critical loads, and 96.5 \% of moderately critical loads, which performs better even with lower efficiencies compared to the battery energy storage systems. To analyze the total served energy, the following resilience index (RI) is used:

\begin{equation}\label{ResilienceIndex}
RI\;(\%)= \Bigg(\frac{Total\;Load - Curtailed\;Load}{Total\;Load}\Bigg) \times 100
\end{equation}

Considering the RI between hours 115 and 144, the worst case scenario is the first one, which is a battery with 2 hours duration and 37.3\% served energy; and the best scenario happened when H2 systems are used as backup power sources and long-duration storages, with an index of 80.1\%.  

\begin{table}[]
\centering
\footnotesize
\captionsetup{labelsep=space,font={footnotesize,sc}}
\caption{ \\ Resilience Analysis Between H2 Systems and Batteries with Different Duration}
\label{RsilienceTable}
\begin{tabular}{|c|c|c|c|c|c|c|}
\hline
\multicolumn{2}{|c|}{Case Study}                                         & 1     & 2     & 3     & 4     & 5     \\ \hline
\multirow{3}{*}{\begin{tabular}[c]{@{}c@{}}\;Energy\;\\ \;Not\;\\ \;Supplied\;\\ \;(MWh)\;\end{tabular}} &
  \begin{tabular}[c]{@{}c@{}}Critical \\ Load\end{tabular} &
  13.92 &
  9.06 &
  4.53 &
  2.11 &
  0 \\ \cline{2-7} 
 & \begin{tabular}[c]{@{}c@{}}\;Moderately\;\\ Critical \\ Load\end{tabular} & 7.57  & 7.57  & 7.24  & 6.03  & 0.38  \\ \cline{2-7} 
 & \begin{tabular}[c]{@{}c@{}}Non-Critical\\ Load\end{tabular}           & 31.38 & 31.38 & 31.38 & 30.17 & 16.41 \\ \hline
\multicolumn{2}{|c|}{\begin{tabular}[c]{@{}c@{}}Total Energy Not \\ Supplied (MWh)\end{tabular}} &
  52.87 &
  48.01 &
  43.15 &
  38.31 &
  16.79 \\ \hline
\multicolumn{2}{|c|}{RI (\%)}                                             & 37.3  & 43.1  & 48.8  & 54.6  & 80.1  \\ \hline
\end{tabular}
\end{table}

\subsection{H2 Production Cost Analysis in Both Normal and Emergency Operation Modes}
The results regarding the H2 production cost including water electrolysis and storage costs are presented in Table \ref{H2cost} for both normal and resilient operation modes. As it can be seen, the capacity factor of electrolyzers in resilient operation mode is higher than normal operation mode due to the DR signal imposed by DSO for preparation of extreme event. Additionally, since H2 demand from FCEVs and the amount of H2 produced for grid assistance are different, capacity factors for three H2 systems are different from each other. Additionally, the average H2 production cost in normal operation is \$1.5/kg. However, in resilient operation mode, the average H2 production cost is \$2.3/kg, due to the power consumption of electrolyzers by expensive DGs. It should be noted that H2 production cost is mainly the function of water electrolysis cost, since it depends on hourly consumed prices rather than the storage cost.

\begin{table}
\centering
\footnotesize
\captionsetup{labelsep=space,font={footnotesize,sc}}
\caption{ \\ H2 Production Cost Analysis in Different Operaton Modes}
\label{H2cost}
\centering
\begin{tabular}{|c|c|c|c|c|c|c|}
\hline
Operation Mode                                                           & \multicolumn{3}{c|}{Normal Operation} & \multicolumn{3}{c|}{Resilient Operation} \\ [0.1 cm] \hline
H2 System                                                                 & HS1        & HS2       & HS3       & HS1         & HS2        & HS3        \\ [0.1 cm] \hline
\begin{tabular}[c]{@{}c@{}}Capacity Factor\\ (\%)\end{tabular}            & 38.34            & 39.27           & 41.73           & 54.74             & 53.29            & 58.11            \\ [0.1 cm] \hline
\begin{tabular}[c]{@{}c@{}}Water Electrolysis\\ Cost (\$/kg)\end{tabular} & 1.46            & 1.54           & 1.59           & 2.18             & 2.33            & 2.53            \\ [0.1 cm] \hline
\begin{tabular}[c]{@{}c@{}}Storage Cost \\ (\$/kg)\end{tabular}           & 0.02            & 0.02            & 0.02           & 0.02             & 0.02            & 0.02            \\ [0.1 cm] \hline
\begin{tabular}[c]{@{}c@{}}H2 Production \\ Cost (\$/kg)\end{tabular}      & 1.48            & 1.57           & 1.61           & 2.21             & 2.35            & 2.55            \\ [0.1 cm] \hline
\end{tabular}
\end{table}

\vspace{-0.2cm}
\section{Conclusion and Future Works}\label{conclusion}
In this paper, a bi-level proactive scheduling framework for H2 systems operation in integrated distribution and transmission networks is proposed for both normal and emergency operation modes. The goal of the paper is to minimize the load curtailment based on their priority and importance using H2 systems. Additionally, rolling horizon approach is used to limit the access of DSO to perfect information regarding the PV, wind, and extreme event time. Additionally, to show the flexibility of H2 systems and preventing from more load curtailment, CBDR signals are modeled for both normal and emergency preparation modes in which DSO asks H2 system owners to fill their storage tank in the case of long-lasting outages (e.g. outages for more than 10 hours). Moreover, realistic costs considering water electrolysis and storage costs are calculated based on the true capacity factor of the electrolyzer. The future work can model load and renewable energies' uncertainty while considering the heat constraints of H2 systems. 

\vspace{-0.2cm}
\section{Appendix}\label{appendix}
The DSO objective function is non-linear (multiplying two continuous variables) due to the $\lambda_{t,i} . P_{t,i}^{Exb}$ term, which is defined as NLE. To address the non-linearity, equation (\ref{dualLL4}) is used to convert it to a linear equivalent equation: 
\begin{equation}
\begin{split}
NLE = & \sum_{t=1}^{T} (\lambda_{t,i} \;.\; P_{t,i}^{Exb}) = \sum_{t=1}^{T}((\underline{\psi}_{t,i} - \rho_{t}^{b}) \;.\; P_{t,i}^{Exb}) = \\ & \sum_{t=1}^{T}( \underline{\psi}_{t,i}\;.\;P_{t,i}^{Exb} - \rho_{t}^{b}\;.\;P_{t,i}^{Exb})   \end{split}  
\end{equation}

Then, by using Karush Kuhn Tucker (KKT) conditions from equation (\ref{exchangebuy}), the following equation is obtained:
\begin{equation}\label{KKT}
\begin{split}
NLE = & \sum_{t=1}^{T}( \underline{\psi}_{t,i}\;.\;P_{t,i}^{Exb} - \rho_{t}^{b}\;.\;P_{t,i}^{Exb}) = \sum_{t=1}^{T} (\underline{\psi}_{t,i}\;.\; P^{UG,max} \\ & - \underline{\psi}_{t,i}\;.\;P^{UG,max}\;.\;U_{i,t} - \rho_{t}^{b}\;.\;P_{t,i}^{Exb}) )
\end{split}  
\end{equation}

Considering equation (\ref{KKT}), the term $\underline{\psi}_{t,i}.P^{UG,max}.U_{i,t}$ is the product of binary and continuous variable, which can be linearized based on the big-M method \cite{wu2016exact} as following:
\begin{equation}
\begin{split}
K_{t,i} = \underline{\psi}_{t,i}.P^{UG,max}
\end{split}  
\end{equation}
\begin{equation}
\begin{split}
-M(1-U_{t,i}) \le K_{t,i} - \underline{\psi}_{t,i}.P^{UG,max} \le M(1-U_{t,i})
\end{split}  
\end{equation}
\begin{equation}
\begin{split}
-M\;.\;U_{t,i} \le K_{t,i} \le M\;.\;U_{t,i}
\end{split}  
\end{equation}

\bibliographystyle{IEEEtran}
\bibliography{mybib}

\end{document}